\documentclass{ws-p9-75x6-50}   

\def\ie{{\it i.e.}\ }
\def\eg{{\it e.g.}\ }

\def\vv{{\it vice versa}}

\def\nonp{non-perturbative}
\def\phi{\varphi}
\def\D{{\cal D}}
\def\p{{p}}
\def\q{{q}}
\def\r{{r}}

\def\x{{\bf x}}
\def\y{{\bf y}}

\def\AA{{\cal A}}
\def\Z{{\cal Z}}

\def\Bb{{\bar B}}

\def\hS{{\hat S}}
\def\e#1{{\rm e}^{\displaystyle #1}}
\def\half{ {\textstyle\frac{1}{2}} }
\def\ker#1{\!\cdot\!#1\!\cdot\!}
\def\ph#1{\phantom{#1}}
\def\str{{\rm str}\,}
\def\tr{{\rm tr}\,}
\def\la{\lambda}

\newcommand{\SU}{\mathop{\rm SU}\nolimits}
\newcommand{\beq}{\begin{equation}}
\newcommand{\eeq}{\end{equation}}

\def\bea{\begin{eqnarray}}
\def\eea{\end{eqnarray}}

\begin{document}

\title{An exact RG formulation of quantum gauge theory}

\author{T.R. Morris}

\address{Department of Physics, University of Southampton,
Highfield, Southampton SO17~1BJ, UK\\E-mail: T.R.Morris@soton.ac.uk}

\maketitle

\abstracts{
A gauge invariant Wilsonian effective action is constructed for pure
$SU(N)$ Yang-Mills theory by formulating the corresponding flow equation.
Manifestly gauge invariant calculations can be performed \ie without gauge
fixing or ghosts. Regularisation is implemented in a novel way which
realises a spontaneously broken $SU(N|N)$ supergauge theory. As an
example we sketch the computation of the one-loop $\beta$ function,
performed for the first time without any gauge fixing.
}

\section{Introduction and motivation}

Our main motivation  is to obtain an elegant
gauge invariant Wilsonian renormalization 
group\cite{kogwil} framework formulated directly in the continuum,
as a first step for non-perturbative analytic approximation
methods.\cite{alg,ymi,ymii} Quite generally such methods can prove
powerful,\footnote{See for example the reviews by Bervillier, Tetradis
and Wetterich in this volume,
and earlier reviews.\cite{YKIS,myrev}}
and of course there is a clear need for a better \nonp\ understanding
of  gauge theory.
However, there are a number of `spin-offs' in solving 
this first step: calculations can be
made without gauge fixing, continuum low energy gauge invariant Wilsonian 
effective actions are for the first time precisely defined,
a four dimensional gauge invariant `physical' regulator is discovered,
and an intimate link to the Migdal-Makeenko equations\cite{mig} is uncovered
(which points to a renormalised version of these Dyson-Schwinger equations
for Wilson loops\cite{polbo}).

We refer the reader to the earlier publications for more detailed
motivation.\cite{alg,ymi,ymii} In this lecture we will concentrate on the
basic steps and try to keep the discussion straightforward and
concrete. The intuition behind these ideas was discussed in an earlier
lecture,\cite{alg} and all the details may be found in the published
papers.\cite{ymi,ymii} However in a number of places, especially for
the more radical steps, we will try to provide some further intuitive
understanding.

In previous exact RG approaches to gauge theory, the authors gauge
fixed, and also allowed the effective cutoff to {break} the gauge
invariance.  They then sought to recover it in the limit that the cutoff
is removed.\cite{typeai} As we have indicated, the present development
follows a very different route. (See also this.\cite{saky})

\section{The Polchinski equation}

We start by casting the established exact RG in a suitable form.
We work in $D$ Euclidean dimensions.
For two functions $f(\x)$ and $g(\y)$ and a
momentum space kernel $W(p^2/\Lambda^2)$, where
$\Lambda$ is the effective cutoff, we introduce the shorthand:
\bea
f\ker{W}g &:=&
\int\!\!\!\!\int\!\!d^D\!x\,d^D\!y\
f(\x)\, W_{\x\y}\,g(\y)\quad, \label{kdefa}\\
{\rm where}\qquad W_{\x\y} &\equiv&\int\!\!{d^D\!p\over(2\pi)^D}
\,W(p^2/\Lambda^2)\,{\rm e}^{i\p.(\x-\y)}\quad.\label{kdefb}
\eea

Polchinski's\cite{Pol} version of Wilson's exact RG,\cite{kogwil}
for the effective interaction
of a scalar field $S^{int}[\phi]$, may then be written
\be
\Lambda{\partial\over\partial\Lambda}S^{int}
=-{1\over\Lambda^2}
 {\delta S\over\delta\phi}^{int}\!\!\!\!\!\ker{c'}
{\delta S\over\delta\phi}^{int}\!\!\!+{1\over\Lambda^2}{\delta
\over\delta\phi}\ker{c'}{\delta S\over\delta\phi}^{int}\quad.
\label{polint}
\ee
Here
$c(p^2/\Lambda^2)>0$ is the effective 
ultra-violet cutoff,
which is implemented by modifying propagators $1/p^2$ to $c/p^2$.
Thus $c(0)=1$ so that low energies are unaltered,
and $c(p^2/\Lambda^2)\to0$ as $p^2/\Lambda^2\to\infty$
sufficiently fast that all Feynman diagrams are ultraviolet regulated.
We may write the regularised kinetic term (\ie the Gaussian fixed point)
as
\be
\hS=\half\,\partial_\mu\phi\ker{c^{-1}}\partial_\mu\phi\quad.
\label{hSs}
\ee
In terms of the total effective action
$S[\phi]=\hS+\,S^{int}$, and $\Sigma_1:=S-2\hS$,
the exact RG equation reads
\be
\label{pol}
\Lambda{\partial\over\partial\Lambda}S
=-{1\over\Lambda^2}
 {\delta S\over\delta\phi}\ker{c'}
{\delta\Sigma_1\over\delta\phi}+{1\over\Lambda^2}{\delta
\over\delta\phi}\ker{c'}{\delta\Sigma_1\over\delta\phi} 
\ee
up to a vacuum energy term that was discarded in (\ref{polint}).\cite{Pol}
(We have more to say on this below.)
The flow in $S$ may be shown directly to correspond to integrating out
higher energy modes,\cite{kogwil,YKIS,wegho,zin,erg,bon}
while leaving the partition function
$\Z= \int\!\!\D\phi\ \e{-S}$ invariant. (For our purposes
we may absorb all source terms into $S$ as spacetime dependent couplings.)
We easily see that $\Z$ is invariant if we rewrite 
(\ref{pol}) as the flow of the measure:
\be
\label{fles}
\Lambda{\partial\over\partial\Lambda}\,\e{-S}=
-{1\over\Lambda^2}{\delta\over\delta\phi}\ker{c'}\left(
{\delta\Sigma_1\over\delta\phi}\,\e{-S}\right)\quad.
\ee
This leaves the partition function invariant 
because the right hand side is a total
functional derivative.

We are about to generalise these ideas in a novel way so it is as well
to settle any nerves about the vacuum energy term we have included in
our version of Polchinski's equation (\ref{pol}).  Of course
 Polchinski was safe in discarding this term from the equations as
uninteresting.  However, his resulting equation then \emph{does not}
leave partition function invariant.  Rather, it evolves with a scale
dependent normalization related to the missing vacuum energy term.  The
extra term we have included is precisely the one discarded and is
precisely the one required to restore the invariance of the partition
function. (As a matter of fact, when flowing with respect to a cutoff
involving only a subset of these fields the included term can even 
become anomalous and crucial to the computation of for example $\beta$
functions.\cite{saky})

\section{Generalisation to gauge theory}
We work with the gauge group $SU(N)$. (All the ideas adapt to other
gauge groups.)
We write all Lie algebra valued quantities as contracted into
the generators. Thus the gauge
field appears as $A_\mu(\x)=A^a_\mu(\x)\tau^a$, the connection for the
covariant derivative $D_\mu=\partial_\mu-iA_\mu$.
Often the coupling $g$ is included in $D_\mu$ but we can choose
to scale it out by absorbing it into $A_\mu$ at the expense
of a non-standard normalisation for its kinetic term.
We will do this for a very important reason, as will become
clear shortly.

The generators $\left(\tau^a\right)^i_{\ph{i}j}$ 
are taken to be Hermitian, in the fundamental representation,
and orthonormalised as ${\rm tr}(\tau^a\tau^b)=\half\delta^{ab}$.
Of course gauge transformations are of the form
$\delta A_\mu=D_\mu\cdot\omega :=[D_\mu,\omega]$
where $\omega(\x)=\omega^a(\x)\tau^a$. 

\begin{figure}[ht]
\epsfxsize=34em 
\epsfbox{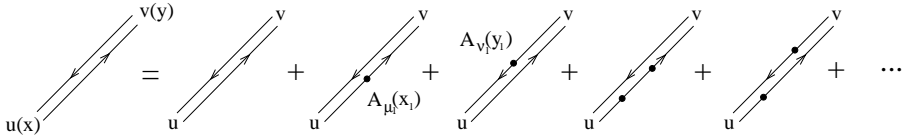} 
\caption{Expansion of the covariantization in terms of gauge fields.
\label{winexp}}
\end{figure}

The question then is how to generalise (\ref{hSs},\ref{pol}) so that the
flow equation is gauge invariant, whilst leaving the partition function
invariant under the flow. It is clear that the regularised
kinetic term must now involve the field strength $F_{\mu\nu}:=i[D_\mu,D_\nu]$,
and some method of covariantizing the cutoff function (which would
otherwise break the gauge invariance). Thus we put
\be
\label{seed}
\hS=\half F_{\mu\nu}\{c^{-1}\}F_{\mu\nu}\quad,
\ee
where the curly brackets is just a short-hand for any given method of
covariantization. To be more explicit about this notation we can write
it in terms of the fundamental indices for any two fields $u$ and $v$
in the $N\otimes{\bar N}$ representation:
\be
\label{wc}
u\{c^{-1}\}v:=\int\!\!d^D\!x\,d^D\!y\,
u^l_i(\x)\,{}^{\ph{x}i}_{\x l}\{c^{-1}\}^k_{j\y}\,v^j_k(\y)\quad.
\ee
Expanding the covariantization in the gauge field $A$ then yields a 
set of vertices (infinite in number except if $c^{-1}$ is a polynomial):
\begin{eqnarray}
u\{c^{-1}\}v&=& 
\sum_{m,n=0}^\infty\int  d^D\!x\,d^D\!y\,
d^D\!x_1\cdots d^D\!x_n\,d^D\!y_1\cdots d^D\!y_m\times
\nonumber\\&&
	\hphantom{\sum_{m,n=0}^\infty\int}\!
\times c^{-1}_{\mu_1\cdots\mu_n,\nu_1\cdots\nu_m}
(x_1,\dots,x_n;y_1,\dots,y_m;x,y) \times
\nonumber\\&&
	\hphantom{\sum_{m,n=0}^\infty\int}\!
\times \tr\Bigl[u(x)\, A_{\mu_1}(x_1)\cdots A_{\mu_n}(x_n)\,
v(y)\, A_{\nu_1}(y_1)\cdots A_{\nu_m}(y_m)\Bigr]\,.\qquad
\label{wv}
\end{eqnarray}
Note that in order to keep the notation compact, we label the resulting 
vertices by the kernel they came from. This procedure can be illustrated
diagrammatically as in fig.~\ref{winexp}.

To be concrete about the covariantization, we could insert
Wilson lines (as suggested by the diagram):
\be
\label{straw}
u\{c^{-1}\}v=\int\!\!\!\!\int\!\!d^D\!x\,d^D\!y\
c^{-1}_{\x\y}\, \tr u(\x)\Phi[{\cal C}_{\x\y}]v(\y)\Phi^{-1}
[{\cal C}_{\x\y}]\quad,
\ee
where $c^{-1}_{\x\y}$ is defined as in (\ref{kdefb}),
${\cal C}_{\x\y}$ is the straight line between $\x$ and $\y$,
and the Wilson line is
the path ordered exponential:
\be
\label{defWl}
\Phi[{\cal C}_{\x\y}]
=P\exp-i\int_{{\cal C}_{\x\y}}\!\!\!\!\!\!dz^\mu
A_\mu(z)\quad.
\ee
More generally we could use curved Wilson lines (and a suitable average
over their shapes and orientations).
Alternatively, we could define the covariantization via the kernels 
momentum representation:
\be
\label{ourchoice}
u\{c^{-1}\}v={\rm tr}\int\!\!d^D\!x\, u(\x)\,c^{-1}(-D^2/\Lambda^2)\cdot v(\x)\quad.
\ee
This is the method we actually used for the calculation of the one-loop
$\beta$ function.

Our notation and method of covariantization can be applied to any kernel.
It is now easy to completely covariantize Polchinski's equation (\ref{pol}):
\be
\label{basRG}
\Lambda{\partial\over\partial\Lambda}S=
-{1\over2\Lambda^2}{\delta S\over\delta A_\mu}\{c'\}
{\delta\Sigma_g\over\delta A_\mu}
+{1\over2\Lambda^2}{\delta\over\delta A_\mu}\{c'\}
{\delta\Sigma_g\over\delta A_\mu}\quad.
\ee
Here and later the $A$ derivatives are defined contracted into the 
generators.\cite{alg,ymi} Just as in (\ref{pol}), 
the first term on the RHS is the {\it classical} term,
yielding the tree corrections, while the second, {\it quantum}, term,
generates the loop corrections. 

We have not yet discussed how the 
coupling $g$ will be incorporated. Recall that we scaled it out of the
covariant derivative. It must appear somewhere in (\ref{basRG}), and with
some thought it turns out that it ends up inside $\Sigma_1$, as\cite{ymi}
\be
\label{sigmag}
\Sigma_g=g^2 S-2\hS\quad.
\ee

By construction, (\ref{basRG}) is manifestly gauge invariant. But we have
also been careful not to disturb the structure that guarantees invariance
of the partition function: indeed, analogously to (\ref{fles}) we have
that the measure still flows by a total derivative:
\be
\label{flesA}
\Lambda{\partial\over\partial\Lambda}\,\e{-S}=
-{1\over\Lambda^2}{\delta\over\delta A_\mu}\{c'\}\left(
{\delta\Sigma_g\over\delta A_\mu}\,\e{-S}\right)\quad.
\ee
This is the crucial property that ensures that lowering $\Lambda$ corresponds
to integrating out. The only other property required is that
momentum integrals really are suppressed above $\Lambda$ -- giving no 
contribution in the limit $\Lambda\to0$. If this is the case, 
since (\ref{flesA}) ensures that $\Z$ is unchanged
under the flow, the contributions from a given fixed 
momentum scale must still be in there somewhere, and the 
only other place they can be, is to already be
encoded in the effective action -- \ie the modes have been
integrated out. 

Of course these equations (\ref{basRG},\ref{sigmag}) are no longer 
equivalent to the Polchinski equation: the covariantizations in
(\ref{basRG}) and (\ref{seed}) lead to many new contributions to $S$
already at tree level. Consistency requires these when we
come to compute physical quantities and \eg the $\beta$ function:
even when there is no gauge fixing something must generate
the eventual contributions normally supplied by ghosts.

The generalised equations (\ref{basRG},\ref{sigmag}) amount to 
considering further scale dependent field redefinitions over and above
those actually implied by the Polchinski equation.\cite{alg,scheme} 
One interpretation
of our equations is that the flexibility
allowed by introducing field redefinitions with each RG `step'
$\Lambda\rightarrow\Lambda-\epsilon$, enables us 
to repair the breaking of the gauge invariance that would otherwise
follow from
just using the Polchinski equation. A deeper interpretation however
is to recognize that the choice of the Polchinski equation is just a 
convenience and not sacrosanct: there are infinitely
many exact RGs just as there are infinitely many ways to block on a 
lattice.\cite{scheme}

Whilst (\ref{sigmag}) shows how the coupling $g$ enters the flow equations,
we have not yet related it to the dynamics of the theory. As usual this
is achieved through a renormalization condition. 

The flow equation is
gauge invariant, Lorentz invariant, and may be Taylor expanded in small 
momenta (for smooth $c$).\cite{ymi} The same is true of the solution
(provided only that a gauge and Lorentz invariant Taylor expandable 
initial bare action is chosen).
Therefore we know that the lowest non-trivial term in a derivative expansion
of the effective action must be $F_{\mu\nu}^2$ up to a so far undetermined
coefficient. This can serve to determine $g$. Thus we write:
\be
\label{deforg}
S={1\over2g^2(\Lambda)}\,{\rm tr}\!\int\!\!d^4\!x\, F_{\mu\nu}^2
+O(\partial^3/\Lambda)
\ee
(ignoring the vacuum energy. To clarify, by $O(\partial^3)$ we mean that
the other gauge invariant terms, each polynomial in derivatives,
would have to contain a part with at least three derivatives; the
full gauge invariant term would of course also contain terms with less
than three derivatives as required by the covariantization.) Note that
in (\ref{deforg}), $g$ occupies the position
that our bare coupling occupies, but in the effective action
we are 
defining the renormalised coupling and anticipate its running
with $\Lambda$. 

Note a very important consequence of the exact preservation of gauge 
invariance: $A_\mu$ has no wavefunction renormalisation. The proof
is so trivial it can take a moment or two to believe it:
if the gauge field were to suffer multiplicative
wavefunction renormalization by $Z$, we would
have to write $A_\mu\mapsto A_\mu/Z$, destroying the gauge invariance since 
then $\delta A_\mu=(Z-1)\partial_\mu\omega+D_\mu\cdot\omega$. This
argument fails in the gauge fixed theory only because $\omega$ is replaced
by a ghost field in the BRST transformation 
leading to pointwise products of fields ($\sim A_\mu\times$
ghost) which are ill defined,
and thus the (BRST) invariance is itself ill 
defined without further renormalization.
Our protection mechanism is familiar from the Background
Field Method,\cite{Abbott} but we stress that here $A_\mu$ is the full
quantum field. 

Note that had we
included the gauge coupling in the covariant derivative as
$D_\mu=\partial_\mu-igA_\mu$, $A$ would then run but oppositely to $g$
in such a way that the actual connection
$gA$ is fixed. Apart from causing more quantities
to run than necessary, it obviously 
also means that the coefficients of an
expansion in $g$ would not be separately gauge invariant.
For these reasons, it is very helpful to scale $g$ out
in the way that we have done.

The result is that (around the Gaussian fixed point) 
the exact preservation of gauge invariance ensures that only
$g$ could receive divergences; $g$ is the
\emph{only} quantity that runs! We see in this observation and
(\ref{deforg}) some small examples of the power and beauty gained by
treating gauge theory in a manifestly gauge invariant manner.  The
simplest na\"\i ve `textbook' argument about $F_{\mu\nu}^2$ being 
the only renormalizable (\ie marginal or relevant) interaction is
immediately true, and not almost lost in the usually rigorously
required complication of ghosts, BRST,
Ward-Takahashi, Lee-Zinn-Justin, and Slavnov-Taylor identities.\cite{zin}

At this point the sceptic might nevertheless wonder whether
gauge fixing is needed in practice. After all, we have only set up
the equations. We have not tried to calculate \eg perturbative amplitudes,
and it is only at this
point that one is forced to gauge fix in the usual approach and 
indeed other initially gauge invariant approaches such as
Migdal-Makeenko equations, and Stochastic quantization.\cite{mig,polbo,zin}
To see explicitly why no problem arises here, we now sketch how
perturbative computations are performed.

\section{Perturbation theory}

With $g^2$ scaled out of the action, it counts powers of Plancks constant.
From (\ref{basRG}) and (\ref{sigmag}) we see that the classical term
is picked out as an order $1/g^2$ piece. Then by iteration we see that
$S$ has the usual weak coupling expansion:
\be
\label{Sloope}
S={1\over g^2} S_0+S_1+g^2 S_2 +\cdots\quad.
\ee
Substituting this back into (\ref{basRG},\ref{sigmag}) 
and recalling that $g$ can
run we find that the beta function takes the expected form
\be
\beta:=\Lambda{\partial g\over\partial\Lambda}=
\beta_1g^3+\beta_2g^5+\cdots\quad,
\ee
with of course so far undetermined coefficients. We will see later how they
are determined. The flow equation (\ref{basRG}) can now be broken up into 
perturbative pieces:
\bea
\label{ergcl}
\Lambda{\partial\over\partial\Lambda}S_0 &=&
-{1\over\Lambda^2}{\delta S_0\over\delta A_\mu}\{c'\}
{\delta\over\delta A_\mu}(S_0-2\hS)\\
\Lambda{\partial\over\partial\Lambda}S_1 &=& 2\beta_1S_0
-{2\over\Lambda^2}{\delta S_1\over\delta A_\mu}\{c'\}
{\delta\over\delta A_\mu}(S_1-\hS)
+{1\over\Lambda^2}{\delta\over\delta A_\mu}\{c'\}
{\delta\over\delta A_\mu}(S_0-2\hS)
\label{ergone}
\eea
and so on. 

\begin{figure}[ht]
\epsfxsize=20em 
$$
\epsfbox{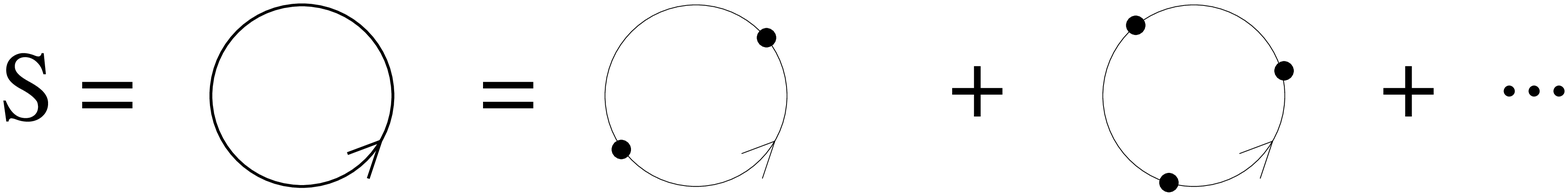} 
$$
\caption{Expansion of the action into traces of gauge fields.
\label{sexp}}
\end{figure}

We can solve these equations by expanding in $A$. By global $SU(N)$
invariance $S$ expands into traces as illustrated in fig.~\ref{sexp}:
\be
\label{Sex}
S =\sum_{n=2}^\infty{1\over n}\int\!\!d^D\!x_1\cdots d^D\!x_n\,
S_{\mu_1\cdots\mu_n}(\x_1,\cdots,\x_n)\
\tr A_{\mu_1}(\x_1)\cdots A_{\mu_n}(\x_n)\ .
\ee
$\hS$ has the same sort of expansion. Actually,
the part of $S$ with four or more gauge fields
also has double trace terms and higher powers of traces,\cite{alg,ymi}
but to simplify matters we ignore them here. 

\begin{figure}[ht]
$$
\Lambda{\partial\over\partial\Lambda}\,
\mathop{\vcenter{\epsfxsize=0.085\hsize\epsfbox{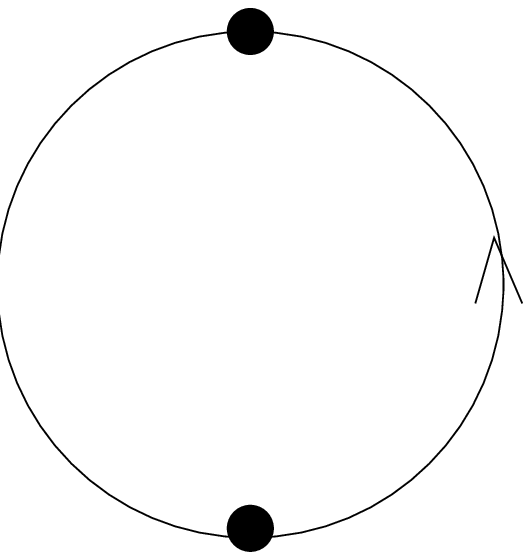}}^{\phantom{p}}
\hskip-.915\hsize}^{\displaystyle
-p^\nu}_{\displaystyle p^\mu}=
\mathop{\vcenter{\epsfxsize=0.13\hsize\epsfbox{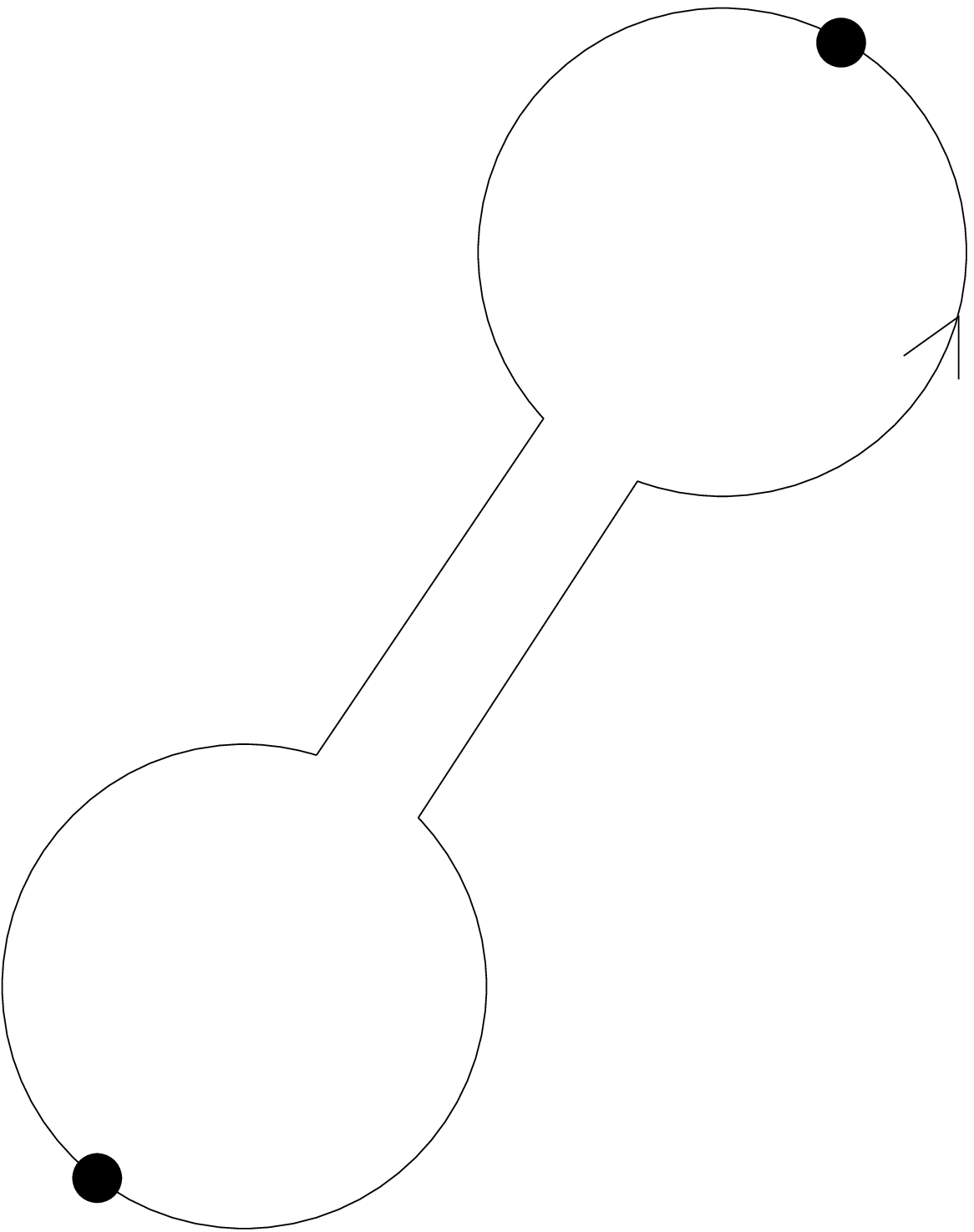}}
\hskip-.87\hsize}_{
\!\!\!\!\!\!\!\!\!\!\!\!\!\!\!\!\displaystyle p^\mu}
^{\displaystyle\;\;\;\;\;\;\;\;\;\;\;\;\;\;\;\;\;\;\;\;-p^\nu}
-2\mathop{\vcenter{\epsfxsize=0.13\hsize\epsfbox{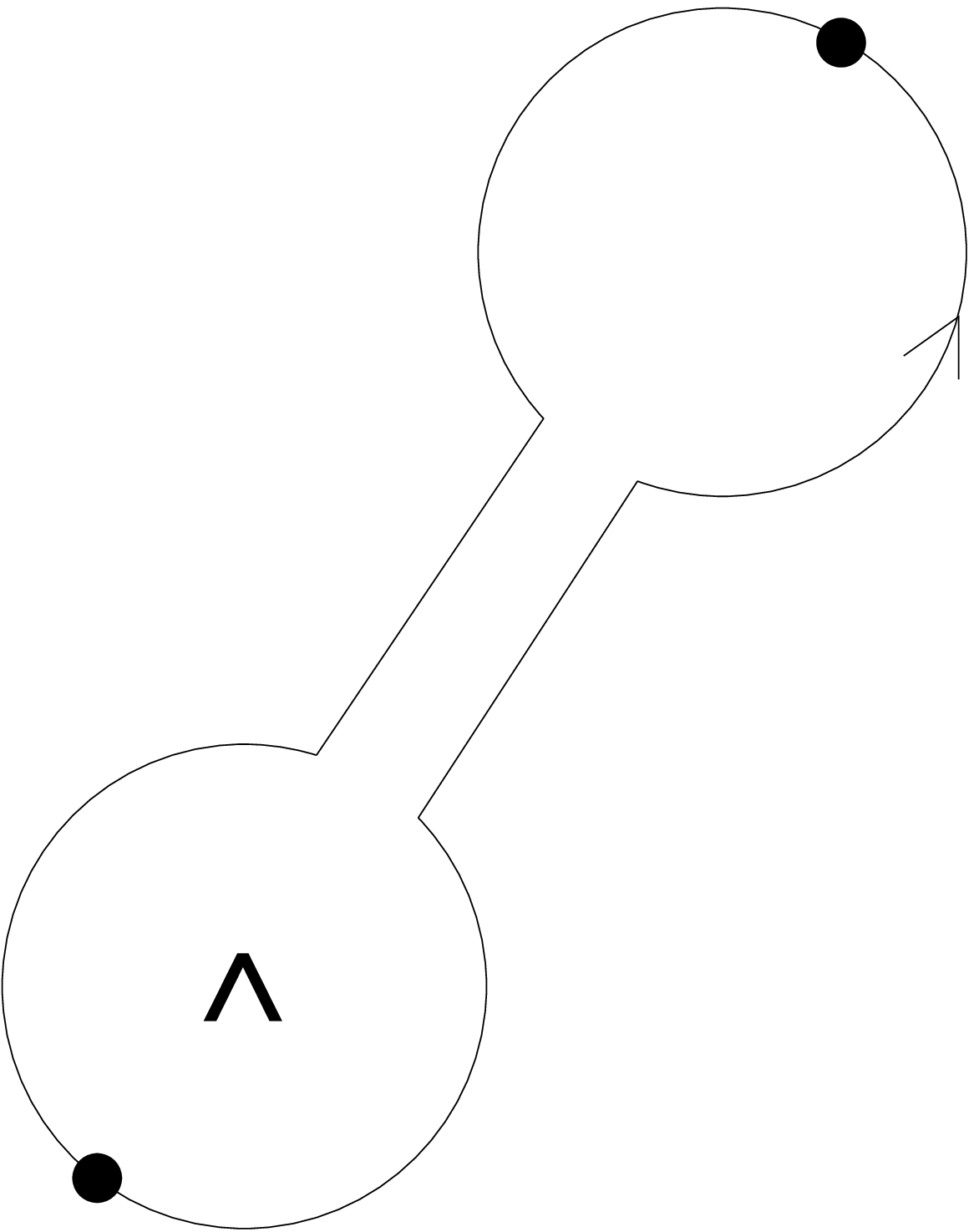}}
\hskip-.87\hsize}_{
\!\!\!\!\!\!\!\!\!\!\!\!\!\!\!\!\displaystyle p^\mu}
^{\displaystyle\;\;\;\;\;\;\;\;\;\;\;\;\;\;\;\;\;\;\;\;-p^\nu}
+(p^\mu\leftrightarrow -p^\nu)$$
\caption{Feynman diagrams for the two-point vertex. Here and later, 
the empty circle corresponds to $S_0$ and the circle with a 
circumflex corresponds to $\hS$.
\label{ftwo}}
\end{figure}

Expanding $S_0$ and $\hS$ as in (\ref{Sex}) we can see that the two-point 
vertex satisfies a flow equation diagrammatically of the form of 
fig.~\ref{ftwo}.
This occurs because the $A$ differentials in (\ref{ergcl}) open up the
traces in fig.~\ref{sexp}, attaching the ends of the
vertices in fig.~\ref{winexp}, as in (\ref{wv}), and thus forming these
gauge invariant dumbell shapes. (Note that since the action has no one-point
vertices, we must place one gauge field in each dumbell, making a 
differentiated two-point vertex. We have not discussed the details of how
the $A$ differentials actually act, 
given that they are also contracted into generators.
This requires using the completeness relation and the simple picture
we have sketched is not quite correct: extra $O(1/N)$ terms
are produced which only contribute to the classical 
six-point vertices and 
higher.\cite{ymi} They will not be needed for this discussion.)

Translating the diagram we have
\be
\label{dtwop}
\Lambda{\partial\over\partial\Lambda}S^0_{\mu\nu}(p)=
-{1\over2\Lambda^2}c'(p^2/\Lambda^2)
\left[S^0_{\mu\lambda}(p)-2\hS_{\mu\lambda}(p)\right]
S^0_{\lambda\nu}(p)+(p_\mu\leftrightarrow -p_\nu)\quad.
\ee
From expanding (\ref{seed}) we readily obtain all the vertices
of $\hS$ in particular
\be
\label{hSex}
\hS_{\mu\nu}(p)=2\Delta_{\mu\nu}(p)/c(p^2/\Lambda^2)\ ,
\ee
where $\Delta_{\mu\nu}(p)=\delta_{\mu\nu}p^2-p_\mu p_\nu$ is the
usual transverse two-point vertex. 
By gauge invariance and dimensions, the solution must take a similar
form:
\be
\label{partwo}
S^0_{\mu\nu}(p)=2\Delta_{\mu\nu}(p)/f(p^2/\Lambda^2)\quad,
\ee
where $f$ is to be determined.
From (\ref{Sloope}), we require $f(0)=1$ so
as to be consistent with (\ref{deforg}) in the $g\to0$ limit. 
Since (\ref{dtwop}) is a first order ordinary differential equation this
boundary condition determines the solution uniquely.
Substituting (\ref{hSex}), we readily find this solution to be $f=c$,
and thus
\be
\label{twop}
S^0_{\mu\nu}(p)=\hS_{\mu\nu}(p)\quad.
\ee

\begin{figure}[ht]
$$
\Lambda{\partial\over\partial\Lambda}\
\vcenter{\epsfxsize=0.085\hsize\epsfbox{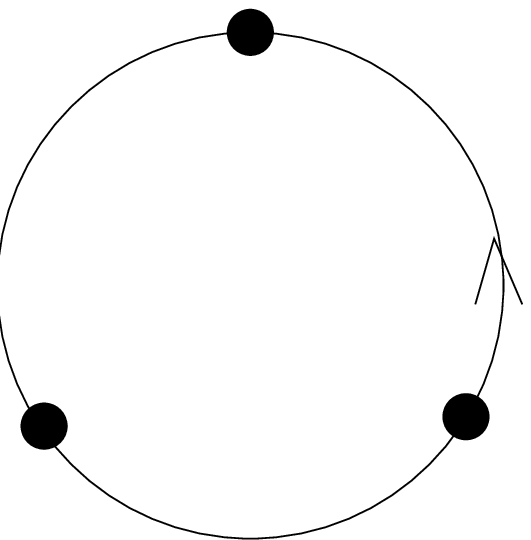}}\hskip-.915\hsize\,
=\, -2\vcenter{\epsfxsize=0.13\hsize\epsfbox{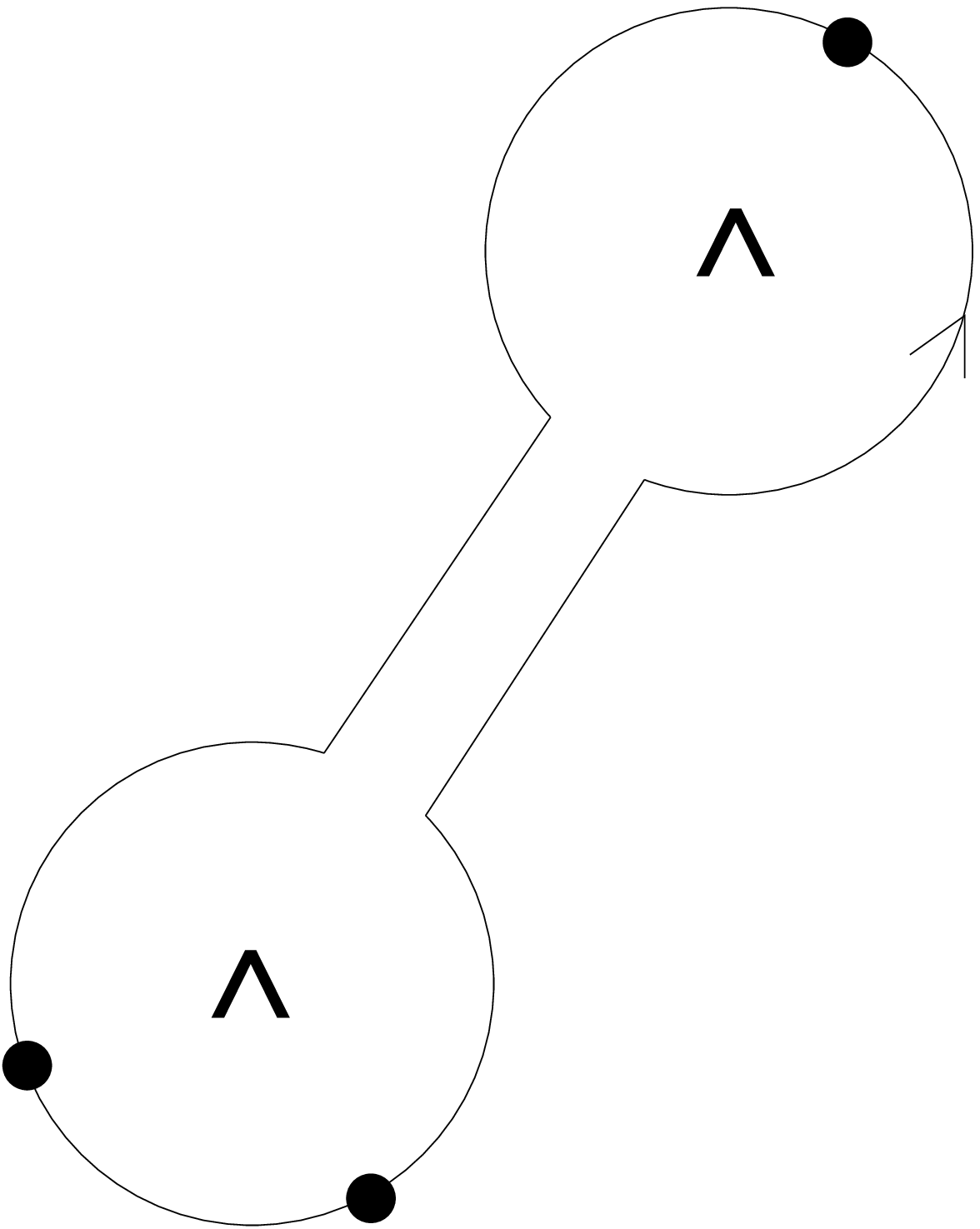}}\hskip-.87\hsize
-2\vcenter{\epsfxsize=0.13\hsize\epsfbox{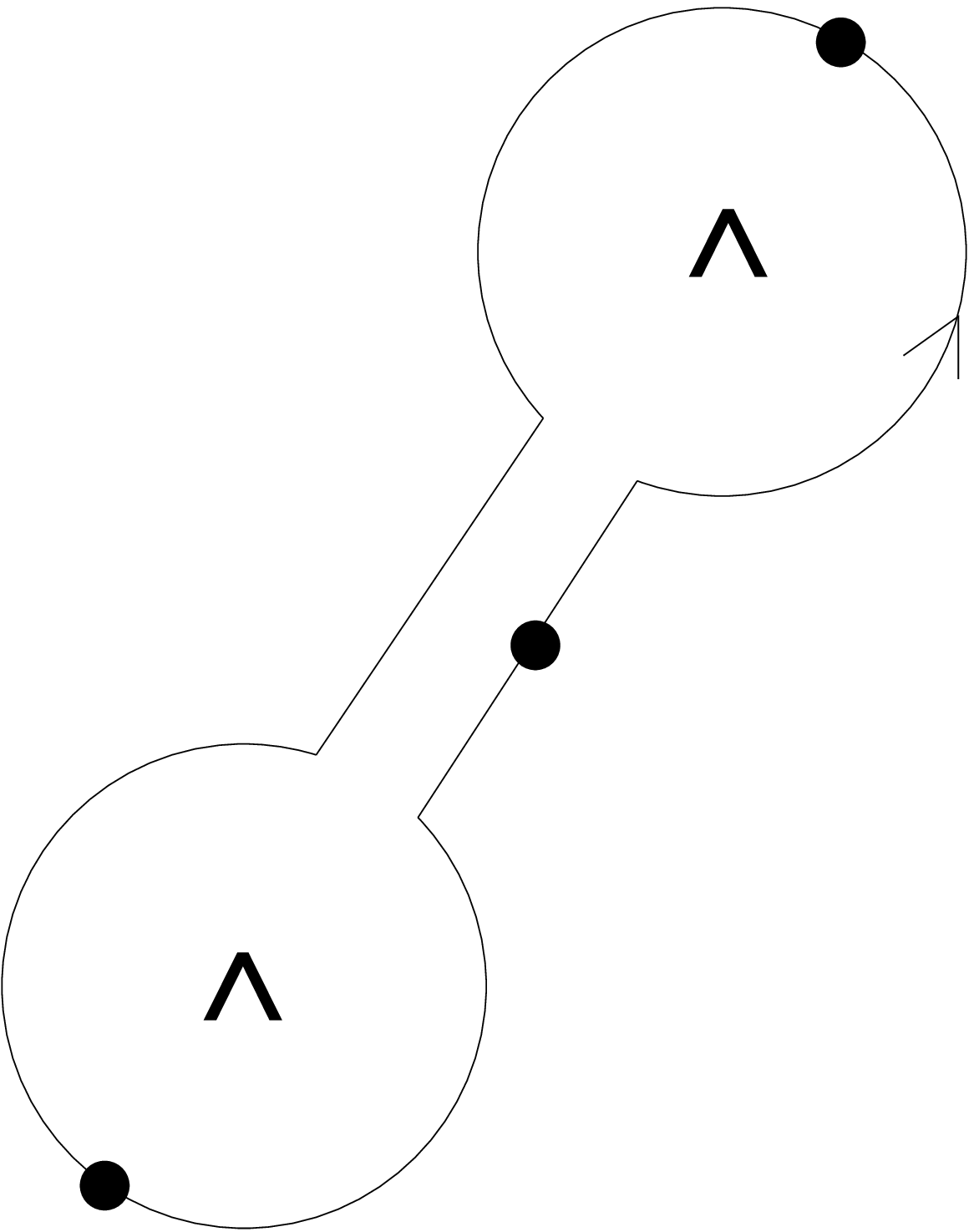}}\hskip-.87\hsize
$$
\caption{Feynman diagrams for the three-point vertex.
The r.h.s. should be summed over cyclic permutations of
the momentum labels. \label{fthree}}
\end{figure}

Higher point vertices fall out even more simply: the fact that the two
two-point vertices agree, results in cancellations so that the right
hand side of the differential equation contains only lower point vertices
that are already determined. This means the result may be immediately
integrated. Let us just illustrate with the three-point vertex. After
the cancellations the diagrams are those of fig.~\ref{fthree},
with solution
\bea
\label{threep}
S^0_{\mu\nu\lambda}(\p,\q,\r) &= &-\int_\Lambda^\infty\!\!{d\Lambda_1\over
\Lambda_1^3}\left\{c'_r
\hS_{\mu\nu\alpha}(\p,\q,\r)\hS_{\alpha\lambda}(r)
+c'_\nu(\q;\p,\r)\hS_{\mu\alpha}(p)\hS_{\alpha\lambda}(r)
\right\} \nonumber \\
&+&2(r_\nu\delta_{\mu\lambda}-r_\mu\delta_{\nu\lambda})\qquad\qquad
+{\rm cycles}\quad. 
\eea
Here it should be understood that in the curly brackets we replace $\Lambda$
with $\Lambda_1$, and to the whole expression we add the two cyclic
permutations of $(p_\mu,q_\nu,r_\lambda)$. We have taken the continuum
limit directly, which is why the top limit is $\infty$ rather than
the overall cutoff $\Lambda_0$. The integration constant is just the usual
bare three-point vertex as follows from (\ref{Sloope}) and the $\Lambda\to
\infty$ limit of (\ref{threep}) and (\ref{deforg}).

In this way we can continue to solve for higher-point classical vertices
and as we will see shortly, also the quantum corrections. {\it Nota
Bene} gauge invariance is maintained at all stages. All the solutions
are gauge invariant. (It is a straightforward exercise\cite{alg,ymi} 
to check that
(\ref{threep}) indeed satisfies the na\"\i ve Ward identity
$p_\mu S^0_{\mu\nu\la}(p,q,r)=S^0_{\nu\la}(r)-S^0_{\nu\la}(q)$.)
At no point are we required to invert some kernel 
that cannot be inverted, which is why gauge fixing is needed in
the usual approach (to form
the propagator).

Now we can explain how the $\beta$ function gets determined. We see
from (\ref{twop}) that 
\be
\label{twopop2}
S^0_{\mu\nu}(p)=2\Delta_{\mu\nu}(p)+O(p^3/\Lambda)\quad.
\ee
Since $F_{\mu\nu}^2$ is the only gauge invariant construct 
that gives this, we know that
the full classical solution has the form
\be
S_0={1\over2}\,{\rm tr}\!\int\!\!d^4\!x\, F_{\mu\nu}^2
+O(\partial^3/\Lambda)
\ee
[with interpretation as in (\ref{deforg})].
Comparing this with (\ref{deforg}) and (\ref{Sloope})
we see that --as usual-- the classical action
\emph{saturates} the renormalisation condition; we learn that 
$S_n$ must have no $F_{\mu\nu}^2$ component for all loops $n\ge1$.
In particular this means that all the loop corrections to the
two-point vertex, \ie $S^n_{\mu\nu}(p)$ $n\ge1$,
must start at $O(p^3)$ or higher. 

Now look at the flow equation for $S^1_{\mu\nu}(p)$,
which can be seen from (\ref{ergone}). The equality (\ref{twop})
kills the middle term on the right hand side. And if we look at only
the $O(p^2)$ contribution the left hand side vanishes, by the arguments
above. This turns (\ref{ergone}) into an algebraic equation and the only
way we can solve it is to choose $\beta_1$ precisely to balance 
(\ref{twopop2}) against the one-loop term (which by gauge invariance is
also proportional to $\Delta_{\mu\nu}$). Thus we have that the $F_{\mu\nu}^2$
part of the one-loop contribution determines $\beta_1$:
\be
\label{beta1}
-{1\over4\Lambda^2}{\delta\over\delta A_\mu}\{c'\}
{\delta\over\delta A_\mu}(S_0-2\hS)\Big|_{F^2_{\mu\nu}}\!\!=\,\,
\beta_1\quad.
\ee
The same is true for the higher coefficients and perturbative flow
equations and even exactly: 
\be
\beta(g)\,=\,-{g^3\over4\Lambda^2}{\delta\over\delta A_\mu}\{c'\}
{\delta\over\delta A_\mu}(S-2\hS)\Big|_{F^2_{\mu\nu}}\!\!\quad.
\ee

\section{Wilson loop interpretation}

Let us mention that we may draw similar diagrammatic representations
to figs.~\ref{ftwo},\ref{fthree} for the full flow equation. We can
use these to show that the large $N$ limit results in $S$ collapsing
to a single trace. (Products of traces decay as
$\sim 1/N$ after appropriate
changes of variables). Of course these diagrams have a close kinship
with `t Hooft's double line notation,\cite{thooft}
but they also have
a deeper meaning in terms of fluctuating Wilson loops. This arises
because the effective action can be expressed as an `average' or
integral over configurations of
Wilson loops (with measure determined by the flow
equation) and the covariantized kernels also have
an interpretation in terms of integrals over Wilson line pairs
[as we have already remarked below (\ref{defWl})]. In this
way we may eliminate the gauge field entirely and reexpress the
flow equation in terms of the natural low energy 
order parameter for a gauge theory: the Wilson loop. In the large
$N$ limit the whole flow equation collapses to a flow for the 
measure over the fluctuations of a single Wilson loop. Describing
its configuration by a particle going round in a circle we see that
we may recast large $N$ $SU(N)$ Yang-Mills as the quantum mechanics
of a single particle (with action determined implicitly through
the flow equation). The details of these ideas may be found
elsewhere.\cite{alg,ymi} Here we keep the discussion firmly
concrete, turning to one-loop calculations.

\section{$SU(N|N)$ regularisation}

\begin{figure}[ht]
$$
2\ \vcenter{\epsfxsize=0.085\hsize\epsfbox{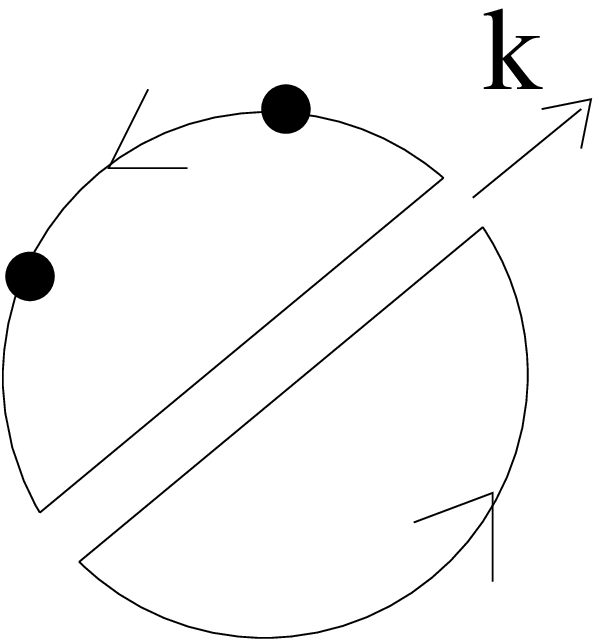}}\hskip-.915\hsize
+2\ \vcenter{\epsfxsize=0.085\hsize\epsfbox{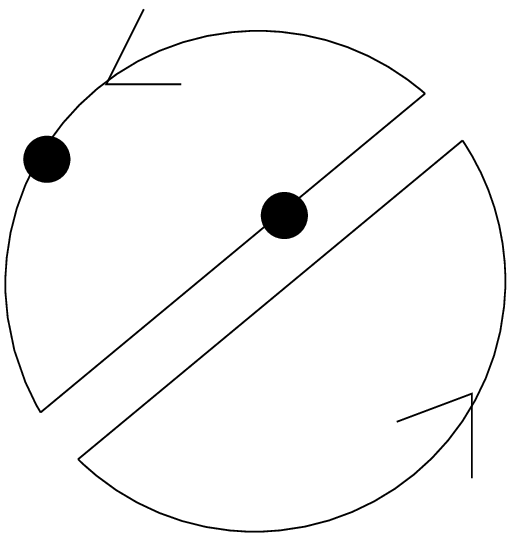}}\hskip-.915\hsize
+2\ \vcenter{\epsfxsize=0.085\hsize\epsfbox{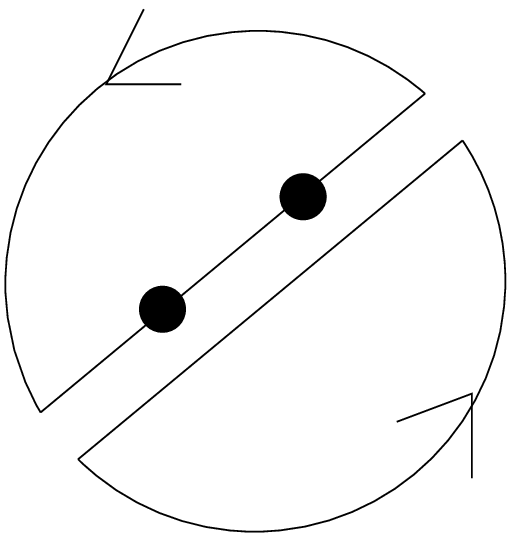}}\hskip-.915\hsize
$$
\caption{One-loop two-point diagrams, constructed from the four-point
vertices of $\Sigma_0=2\hS-S_0$. \label{foneloop}}
\end{figure}

From (\ref{beta1}), the diagrams contributing to $\beta_1$ are
those of fig.~\ref{foneloop}. Here the trace has been broken
open in two places and rejoined by the kernel to form two 
traces.\footnote{Again we here ignore some terms arising from the completeness
relation, which happen to vanish.\cite{ymi}}
The lower one being empty, just contributes $\tr1=N$. 

Once
again, it is a simple matter to translate the diagrams to algebra.
This time of course we have a momentum integral to do. When we try
to compute it however we find the result diverges. This is no
surprise: by using covariantized cutoff functions we have effectively
implemented covariant higher derivative regularisation, which is
known to fail.\cite{oneslip} Actually it cures the superficial divergences 
of all higher
loops except one loop, but of course superficial is not enough. 
In standard perturbation theory, despite some early 
controversy\cite{pv,warr} this problem has been cured by
supplementing the higher derivative regularisation with a system of
Pauli-Villars regulator (PV) fields, the action being bilinear in
these fields so that they provide, on integrating out, the missing one
loop counterterms.\footnote{And of course other finite
contributions.}\cite{slavnov}  This solution turns out to be
unwieldy, but worse, here the property of being bilinear in the PV
fields is not preserved by the flow:\cite{alg} 
as the gauge field is integrated
out higher-point PV interactions are generated.

Instead, we uncovered a system of regulating fields that is more
natural from the exact RG point of view.\cite{alg,ymi,ymii}  We
have gradually realised that hidden in this formulation are
supermatrices and a spontaneously broken local $\SU(N|N)$. 
Whilst many aspects fell
out correctly without us being aware of this structure, 
the formulation we developed amounts to a unitary gauge in the fermionic
directions and
is limited to one loop.\cite{alg,ymii}  
Complete regularisation is achieved in
a fully local $\SU(N|N)$ framework as will be seen in
the next lectures.\cite{sunn} 
(Needless to say,
such a framework may be used independently of the Wilsonian
RG, and provides a novel and elegant four dimensional
`physical'\footnote{In the usual sense that it directly suppresses
higher momentum modes.} regularisation for gauge theory.)

Let us stress that there are \emph{two} main threads
here. On the one hand we introduce this natural gauge invariant
regularisation, as described above. On the other hand, we go on to use
it to repair the divergences in the gauge invariant exact RG flow 
equation, and thus
develop a consistent calculational framework in which manifest gauge
invariance can be maintained at all stages.  

Initially we discovered the regularisation intuitively 
from the bottom up, introducing interactions in such a way as
to guarantee that divergences in one diagram would be cancelled
by another at any stage of the flow. Iterating this procedure
it turns out that there is very little freedom,
and essentially a unique set of Feynman rules for the
PV fields is found.\cite{alg} 

To get a flavour of this, let
us just describe how the spectrum of PV fields is determined.
First we note that for these high momentum contributions to cancel
pairwise amongst diagrams, we need the
PV field to have, at least at high momentum, the \emph{same} vertices
as $A_\mu$. Therefore we introduce an adjoint field $B_\mu$.
In order for it to contribute the opposite sign in
one-loop contributions, $B$ must be fermionic. However, when contracted
into the Bose-symmetric vertices from $A$, its anticommutation properties
will cause (parts of) the result to vanish. 
We cure this problem by making it complex,
so that we can `pepper' the $A$-vertices with $B_\mu$ and $\Bb_\mu$.
However, our problems are not over because we then have many more
vertices with $B$ and $\Bb$ in, than with just $A$, leading to many
divergent Feynman diagrams with no cancelling partner. 
We can cure this with the following idea.
We double the
gauge group to $SU(N)\times SU(N)$. The original gauge
field will be written $A_\mu^1$. We introduce $A^2_\mu$
belonging to the second $SU(N)$. We place
$B^{i_1}_{j_2}$ in the `middle',
fundamental with respect to $SU_1(N)$ and complex conjugate fundamental
in $SU_2(N)$ (thus oppositely for $\Bb^{i_2}_{j_1}$). In this way
group theory constrains $B$ to
follow $\Bb$ and \vv, when tracing round a vertex,
restoring the pairwise identification of gauge field diagrams
with PV diagrams. Finally, extra divergences arise from the fact that
the $B$s are massive and thus have longitudinal components. These
can be cancelled by introducing bosonic scalars $C^i$ with (covariant)
derivative interactions. 

We now have a much more elegant way of arriving at this:\cite{ymii,sunn}
we extend $SU(N)$ Yang-Mills to one built on the 
supergroup $SU(N|N)$.\cite{bars}
The gauge field becomes a supergauge field:
\begin{equation}
\AA_\mu =\pmatrix{A^1_\mu&B_\mu\cr \Bb_\mu&A^2_\mu}\quad,
\label{superg}
\end{equation}
Gauge invariance extends to a full supergauge invariance:
$\delta\AA_\mu = \nabla_\mu\cdot\omega$,
where $\nabla_\mu=\partial_\mu-i\AA_\mu$ is the supercovariant derivative
and $\omega$ is now also a supermatrix. Covariantization of the
flow equation under this new group is straightforward. For example,
$\hS$ becomes 
\be
\hS=\half{\cal F}_{\mu\nu}\{c^{-1}\}{\cal F}_{\mu\nu}\quad,
\ee
where ${\cal F}_{\mu\nu}$ is the superfield strength, covariantization
in (\ref{wv}) is via $\AA$ and, most importantly, all traces here and
elsewhere are replaced by supertraces. This last step is necessary to
preserve the property of cyclicity when using supermatrices (which in
turn is necessary to prove gauge invariance). The supertrace is defined
by
\be
\str {\bf X} =\str\pmatrix{X^{11} &X^{12}\cr X^{21} &X^{22}}=\tr
X^{11}-\tr X^{22}\quad.
\ee
This means that in the quantum corrections (such as the $\beta_1$
calculation above) the $\tr1=N$ parts are now replaced by
$\str1=0$ ! The symmetry between bosonic and fermionic 
contributions causes quantum corrections to vanish, just as it can
with normal space-time supersymmetry. (Here however the supersymmetry
is implemented in a novel way: on the fibre.) 

We are not interested
in such a complete cancellation but then we are not interested in massless
fermionic vector fields $B_\mu$ either. We must make these fields massive
without destroying the cancellation properties at high energies (which 
will then act as a regulator) and without disturbing the original
$SU(N)$ gauge invariance. Fortunately we know how to do this: we
introduce a superscalar `Higgs' to spontaneously break the theory
along just the fermionic directions, \ie
\be
{\cal C}=\pmatrix{C^1&D\cr {\bar D}&C^2}\qquad{\rm such\ that}\qquad
<\!{\cal C}\!>=\Lambda\pmatrix{1&0\cr0&-1}\quad.
\ee
In unitary gauge $B$ eats $D$ and becomes massive of order $\Lambda$, 
leaving behind the `physical' Higgs
$$
\pmatrix{C^1&0\cr 0&C^2}\quad,
$$
whose mass is also (or may naturally be chosen to be) of order $\Lambda$,
and an undisturbed $SU(N)\times SU(N)$ gauge invariance.
Remarkably, this construction 
leads to the \emph{same} spectrum and interactions as the bottom up
approach!

There are a number of subtleties
that we have skated over: we need to higher-derivative regularise
also the Higgs sector and  this introduces another cutoff function,
which here for simplicity we have ignored.\cite{ymii,sunn} 
Of course $A^1$ remains massless, but the second gauge
field $A^2$ also remains massless, and is unphysical since it 
has a wrong sign action\cite{alg} (a 
consequence of the supertrace\cite{ymii}). This leads to a source
of unitarity breaking which however disappears in the limit the cutoff
is removed.\cite{sunn} (In the exact RG approach this corresponds to
setting $\Lambda_0\to\infty$.) This is because $A^1$ and $A^2$ are the
gauge fields of the direct product subgroup $SU(N)\times SU(N)$, and 
are therefore charge neutral under each others
gauge group, thus at energies much less than $\Lambda_0$, 
we are left only with these gauge fields which decouple. To prove this
it is only necessary to show that the
amplitudes that mix the two $SU(N)$s all vanish at fixed momentum
as $\Lambda_0\to\infty$. This does indeed follow, from gauge invariance
and dimensional considerations.\cite{sunn} Finally, 
while quantum corrections are now finite, this results from cancellation
of separately
divergent pieces and thus care is needed in defining these conditionally
convergent integrals
(\eg by employing a gauge invariant preregularisation\cite{alg,ymi,ymii}).

Recomputing the one loop $\beta$ function with the manifestly gauge
invariant exact RG extended to this realisation of
spontaneously broken $SU(N|N)$, we
find that the momentum integral is now finite and furthermore the
integrand is a total divergence. The resulting surface integral is
independent of the choice of cutoff function $c(p^2/\Lambda^2)$ and
treating the implied limits carefully yields the famous result\cite{ymii}
$$\beta_1=-{11\over3}{N\over(4\pi)^2}\quad,$$
thus furnishing the first calculation of the one-loop $\beta$
function without fixing the gauge.

\section*{Acknowledgments}
TRM wishes to thank PPARC for financial support through visitor, SPG
and Rolling grants
PPA/V/S/1998/00907, PPA/G/S/1998/00527 and GR/K55738.

\end{document}